\documentclass[runningheads]{llncs} 
\usepackage{hyperref}
\usepackage{latexsym,amssymb} 
\usepackage{color} 
\usepackage{listings} 
\usepackage{mathpartir} 
\usepackage[only,Mapsto]{stmaryrd} 
\usepackage{relsize}

\newif\iflongver\longvertrue 

\iflongver
\usepackage{endnotes}
\usepackage{myhyperendnotes} 
\newcommand{\longversion}[1]{#1} 
\newcommand{\shortversion}[1]{} 
\newcommand{\myendnote}[1]{\endnote{#1}} 
\else
\newcommand{\longversion}[1]{}
\newcommand{\shortversion}[1]{#1} 
\newcommand{\myendnote}[1]{}
\fi

\newcommand{\dt}[1]{\textbf{\emph{#1}}} 

\newcommand{\sep}{\mathbin{\mid}} 
\newcommand{\Sep}{\mathbin{\:\mid\:}} 

\newcommand{\choice}{+} 

\newcommand{\converse}{^{\sim}} 
\newcommand{\prefix}{\preceq} 

\renewcommand{\neg}{\mbox{\relsize{-1}$\lnot$}}

\renewcommand{\phi}{\varphi}
\newcommand{\ol}[1]{\overline{#1}}
\newcommand{\union}{\cup}
\newcommand{\intersect}{\cap}
\newcommand{\proves}{\vdash}
\newcommand{\aftqua}{.\:}
\newcommand{\all}[2]{\forall #1 \aftqua #2}
\newcommand{\some}[2]{\exists #1 \aftqua #2}

\newcommand{\imp}{\Rightarrow}

\newcommand{\rt}{\sqsubseteq}

\newcommand{\ctrl}{\mathconst{ctrl}}

\newcommand{\id}{\mathconst{id}}
\newcommand{\Left}{\mathconst{left}}
\newcommand{\Right}{\mathconst{right}}
\newcommand{\destutter}{\mathconst{destutter}}

\newcommand{\last}[1]{#1_{-1}} 
\newcommand{\fin}{\mathit{fin}}
\newcommand{\trans}{\mapsto}
\newcommand{\biTrans}{\Mapsto} 

\newcommand{\uequiv}{\mathrel{\cong}} 

\newcommand{\mathconst}[1]{\mbox{\textsl{#1}}} 
\newcommand{\rn}[1]{\textsc{\relsize{-1}#1}} 

\newcommand{\leftex}[1]{ \raisebox{.25ex}{\relsize{-1}$\langle\hspace*{-2.1pt}[$} #1 \raisebox{.25ex}{\relsize{-1}$\langle\hspace*{-2.0pt}]$} }
\newcommand{\rightex}[1]{ \raisebox{.25ex}{\relsize{-1}$[\hspace*{-2.0pt}\rangle$} #1 \raisebox{.25ex}{\relsize{-1}$]\hspace*{-2.1pt}\rangle$} }

\newcommand{\leftF}[1]{\leftex{#1}}
\newcommand{\rightF}[1]{\rightex{#1}}
\newcommand{\Both}[1]{\ensuremath{\mathbb{B}} #1}
\newcommand{\Agr}{\ensuremath{\mathbb{A}}}

\newcommand{\keyw}[1]{\ensuremath{\mathsf{#1}}}

\newcommand{\skipc}{\keyw{skip}}
\newcommand{\ifc}[3]{\keyw{if}\ {#1}\ \keyw{then}\ {#2}\ \keyw{else}\ {#3}}
\newcommand{\whilec}[2]{\keyw{while}\ {#1}\ \keyw{do}\ {#2}\ \keyw{od}}
\newcommand{\varblock}[2]{\keyw{var}~ #1 ~\keyw{in}~ #2} 

\renewcommand{\P}{\mathcal{P}}  
\renewcommand{\S}{\mathcal{S}} 
\newcommand{\Lrel}{\mathcal{L}} 
\newcommand{\Q}{\mathcal{Q}} 
\newcommand{\R}{\mathcal{R}} 
\newcommand{\I}{\mathcal{I}} 
\newcommand{\E}{\mathcal{E}} 
\newcommand{\F}{\mathcal{F}}

\newcommand{\subst}[3]{{#1}^{#2}_{#3}}

\newcommand{\specSym}{\leadsto}
\newcommand{\rspecSym}{\ensuremath{\mathrel{\mbox{\footnotesize$\thickapprox\hspace*{-.4ex}>$}}}}

\newcommand{\spec}[2]{#1\specSym #2} 
\newcommand{\rspec}[2]{#1\rspecSym #2} 

\newcommand{\fspec}[3]{#1\specSym #2\:[#3]} 

\definecolor{dkgreen}{rgb}{0,0.6,0}

\lstdefinelanguage{TOY}
{
basicstyle=\normalsize\rmfamily\itshape, 
keywordstyle=[1]\upshape\color{blue}\bf,%
morekeywords=[1]{while,do,od,%
var,if,then,else,fi,skip%
},%
string=[b]",%
commentstyle=\itshape\color{dkgreen},%
columns=[l]fullflexible,%
sensitive=true,%
morecomment=[s]{(*}{*)},%
numberstyle=\upshape,
keepspaces=true,%
literate=%
 {\\expx}{{$2^x\:$}}1  
 {\\uequiv}{{$\:\uequiv\:$}}1  
 {*<}{$\lexop$}{1}
 {*]}{$\lexcl$}{1}%
 {*>}{$\rexcl$}{1}%
 {*[}{$\rexop$}{1}%
 {=:=}{$\eqbi$}{1}%
 {<}{$<$}{1}%
 {>}{$>$}{1}%
 {-}{$\,-$}{1}%
 {'}{$\,^\prime$}{1}%
 {<=}{$\le$}{1}%
 {>=}{$\ge$}{1}%
 {<>}{$\ne$}{1}%
 {/\\}{$\land$}{1}%
 {mod}{{\upshape mod}}{1}%
 {\\/}{ $\lor$ }{3}%
 {not\ }{$\lnot$ }{2}%
}

\titlerunning{Thirty-seven years of relational Hoare logic}
\title{Thirty-seven years of relational Hoare logic:
remarks on its principles and history%
\longversion{\protect\\(extended version)}
}
\author{David A. Naumann\longversion{\thanks{Partially supported by NSF CNS 1718713 and 
ONR N00014-17-1-2787.}} 
}
\institute{Stevens Institute of Technology
}
\begin{document}
\maketitle


\begin{abstract}
Relational Hoare logics extend the applicability of modular, deductive verification to encompass important 2-run properties including dependency requirements such as confidentiality and program relations such as equivalence or similarity between program versions.  A considerable number of recent works introduce different relational Hoare logics without yet converging on a core set of proof rules. This paper looks backwards to little known early work. This brings to light some principles that clarify and organize the rules as well as suggesting a new rule and a new notion of completeness.
\end{abstract}


\longversion{{\color{red}%
A shorter version of this paper appears in ISOLA,
the 9th International Symposium On Leveraging Applications of Formal Methods, Verification and Validation (Springer LNCS 12477, pages 93--116).
This version has additional material as endnotes, and minor revisions.
}}

\section{Introduction}\label{sec:intro}


Even in the archivally published part of the scientific literature,
there are some gems known to few but deserving the attention of many.  
Such a gem is a paper by Nissim Francez published in 1983,
around the time of Apt's two-part paper ``Ten Years of Hoare Logic''~\cite{Apt10a,Apt10b}.
Relational Hoare Logic (RHL) formalizes reasoning about two programs. 
The term, and a version of the logic, are introduced in a well known gem by Nick
Benton published in 2004~\cite{Benton:popl04}.
Relating two programs is far from new, and is important: it encompasses equivalence (as in compilation),  refinement and conditional similarity (as in software development, 
evolution, and re-engineering), and properties of a single program
(like determinacy of output) for which one must consider two executions.
Reasoning about two executions inevitably leads to reasoning about two different programs---that is one of the principles already articulated in the paper by Francez
titled ``Product Properties and Their Direct Verification''~\cite{Francez83} 
which introduces many rules of RHL.  

The fundamental safety property is \emph{partial correctness}: for each of the program's runs,
if the initial state satisfies the designated precondition, and the run terminates, then the final state satisfies the designated postcondition.
The fundamental liveness property is \emph{termination}: for each of the program's runs,
if the initial state satisfies the precondition then the run is finite.
Many interesting or desirable behavioral properties of a program are such \emph{trace properties}, that is,
a condition on runs is required to hold for all runs.
Relations between programs involve two runs at a time, for example one notion of equivalence is that from the same initial state, runs of the two programs reach the same final state.  
One cannot expect this property if the programs are nondeterministic.  
What then is determinacy? It is the property of a program that from any initial state, its runs all diverge or all terminate in the same state.
This can be defined more parsimoniously: from any initial state, any two runs either both diverge or both terminate in the same state.  
Behavioral program properties involving multiple runs have been dubbed \emph{hyperproperties} and 
an important subclass are the $k$-\emph{safety} properties which, for some fixed $k\geq 1$ can be defined  
by requiring all $k$-tuples of runs satisfy a condition on $k$-tuples~\cite{TerauchiAiken,ClarksonSchneiderHyper10}.  
Francez uses the term \emph{power property} for what is now called $k$-safety, 
and \emph{product property} for relations between programs.

The $k$-safety properties are an attractive object of study because they are amenable to reasoning techniques 
that are natural generalizations of those for safety properties.
As is likely to occur to anyone familiar with programming or automata theory,
to prove a 2-safety property one may construct a \emph{product program} whose runs represent 
pairs of runs of the original.  
Francez points out that product programs can be expressed as ordinary programs,
so that Hoare logic (HL) can be used to prove 2-safety:
If $C$ is a sequential program acting on variables, we can choose fresh names and obtain 
a renamed copy $C'$, and then $C;C'$ serves as a product.
This particular product construction is often called \emph{self-composition}, a term from Barthe et al~\cite{BartheDArgenioRezk,BartheDR11} who rediscover the idea (and also consider other forms of product)
for proving information flow security.

By now, scientific interest together with practical importance has led to exciting achievements. 
Related challenges are under active study by a number of research groups, 
often making use of some form of RHL.
Quite a few papers have appeared with RHLs,
some emphasizing frameworks meant to provide unifying principles~\cite{Benton:popl04,BartheCK16,AguirreBGGS19,MaillardHRM20},
but there is considerable variety in the proof rules included or omitted.
By contrast, the core rules of HL for imperative programs appear in many places with relatively little variation.
There is a scientific explanation for this situation: the recipe for boiling a logic down to its essentials
is to state and prove a completeness theorem that says true properties are provable.
But, through product programs, relational reasoning is reduced to HL---so completeness in this sense  
is a trivial consequence of completeness for \emph{unary} (i.e., 1-safety) properties,
as Francez observes.  His paper concludes with a problem that is still open:
``It would be interesting to obtain a formal characterization of the situation in which the proposed method achieves actual proof simplification over the indirect proofs using Hoare's (unary) logic.''

In this paper I describe, in as plain terms as I can, various reasoning principles and 
their embodiment in proof rules.
One contribution is to systematize knowledge 
and in particular to emphasize the importance of 
program equivalences to reduce the number of core rules.
I also introduce a new rule to fill a gap that becomes evident when one considers
one of the key motivations for relational reasoning.
Finally, I introduce a new notion:  alignment completeness.
It appears naturally when one recapitulates, as Francez does and I do with slightly more generality, 
the development from Floyd to Hoare.

Scientists thrive on getting credit and good scientists take care in giving credit.  But it is not always easy to determine the origin of ideas, in part because good ideas may be independently rediscovered several times before becoming firmly ensconced in the edifice of the known.  
My primary aim in this paper is to explain some principles as I have understood them, 
not to give a survey of prior work. 
I do point out ideas found in the paper by Francez, and cite some other work in passing.
Other early works that may have been overlooked can be found in the first paragraph of Sec.~\ref{sec:related}.

\paragraph{Outline.}
Following background on the inductive assertion method and HL (Sec.~\ref{sec:prelim}),
the method is extended to aligned pairs of runs in Sec.~\ref{sec:align},
as background for RHL which comprises Sec.~\ref{sec:RHL}.
Sec.~\ref{sec:related} discusses related work and Sec.~\ref{sec:concl} concludes.
\shortversion{An extended version of the paper is at \url{http://arxiv.org/abs/1910.14560}.}
\longversion{\emph{In this long version of the paper, superscript numerals refer to end notes.
These provide additional details including related work on completeness.
}}

\section{Preliminaries}\label{sec:prelim}

\subsection{The inductive assertion method}

We focus on the simple imperative or ``while'' language
with assignments and possibly other primitive commands like nondeterministic assignment.  
The reader is expected to be familiar with transition semantics, 
in which the program acts on stores, where a store is a total mapping from variables to values.\myendnote{Consult, for example, the book ~\cite{AptOld3}.
  There may be infinitely many variables but any program has the \dt{frame property}:
  there is a finite set of variables (or memory locations) such that the program acts on, and depends on, only 
  those, leaving the rest unchanged.}  
The following abstraction of a program's semantics is convenient.

An \dt{automaton} is a tuple $(Ctrl,Sto,init,\fin,\trans)$ where 
$Sto$ is a set (the data stores),
$Ctrl$ is a finite set 
that contains distinct elements $init$ and $\fin$, 
and ${\trans} \subseteq (Ctrl\times Sto)\times(Ctrl\times Sto)$ is the transition relation.
We require 
$(c,s)\trans (d,t)$ to imply $c\neq \fin$ and $c\neq d$ 
and call these the \dt{finality} and \dt{non-stuttering} conditions respectively.\myendnote{This loses no generality and facilitates the definition, later, of $destutter$.}
A pair $(c,s)$ is called a \dt{state}.  Let $\beta$ and $\gamma$ range over states.
A \dt{trace} of an automaton is a non-empty sequence $\tau$ of states, consecutive under the transition relation, with $\ctrl(\tau_0)=init$.
It is \dt{terminated} provided $\tau$ is finite and $\ctrl(\last{\tau})=\fin$,
where $\last{\tau}$ denotes the last state of $\tau$. 

In structural operational semantics, transitions act on configurations of the form $(c,s)$ where $c$ is a command, and $\skipc$ by itself indicates termination.
This fits our model: take $init$ to be the program of interest, $Ctrl$ to be all commands,\myendnote{For
   some variations of structural operational semantics it suffices for $Ctrl$ to be something like
   finite sequences of subprograms of the program of interest.}
and $\fin$ to be $\skipc$.
Another instantiation treats $Ctrl$ as the points in the program's control flow graph (CFG).

A partial correctness property is given by a pair of store predicates, $P,Q$, for which we write
$\spec{P}{Q}$.
In a formal logic, $P$ and $Q$ range over formulas in some assertion language, usually first order logic for a designated signature that includes types and operations used in the program.
We write $s\models P$ to say $s$ satisfies predicate $P$, and define $(c,s)\models P$ iff $s\models P$.
As a means to specify requirements, the notation is inadequate.  
The postcondition $y=x+1$ can be achieved by changing $x$ or by changing $y$.  
This problem is best solved by including a \dt{frame condition} which for simple imperative programs is just a list $\ol{x}$ of variables permitted to change, which we write as $\fspec{P}{Q}{\ol{x}}$.
Its meaning can be reduced to the simpler form provided we distinguish between program variables and spec-only variables not allowed to occur in programs.\myendnote{Because then we can write
  $\spec{\hat{P}}{\hat{Q}}$ where $\hat{P}$ conjoins to $P$ some equations of the form $x=\hat{x}$,
  one for each program variable $y$ not in $\ol{x}$, with each  $\hat{x}$ a distinct spec-only variable
  not free in $P$ or $Q$. Conjoining the same equations to $Q$ gives the property that program variables
  $y\notin\ol{x}$ are unchanged, given the stipulation that spec-only variables are never changed.}
That being so, we focus on the form $\spec{P}{Q}$ and use for it the succinct term \dt{spec}.

Let us spell out two semantics for specs in terms of an automaton $A$.
The \dt{basic semantics} is as follows.
For a finite trace $\tau$ to satisfy $\spec{P}{Q}$ means that $\tau_0\models P$ 
and $ctrl(\last{\tau})=\fin$ imply $\last{\tau}\models Q$,
in which case we write $\tau\models \spec{P}{Q}$.
Then $A$ satisfies $\spec{P}{Q}$ just if all its finite traces do.
The \dt{non-stuck semantics} adds a second condition for $\tau$ to satisfy the spec:
$ctrl(\last{\tau})\neq \fin$ implies $\last{\tau}\trans -$,
where $\last{\tau}\trans -$ means there is at least one successor state.
Stuck states are often used to model runtime faults.

The inductive assertion method (IAM) of Floyd~\cite{Floyd67short},\myendnote{A detailed 
   history of these ideas and terminology, which go back to Turing~\cite{TuringChecking49}, is 
   provided by Apt and Olderog in their article on Hoare's logic~\cite{AptO19}.}
is a way to establish that command $C$ satisfies spec $\spec{P}{Q}$.
The first idea is to generalize the problem: in addition to establishing that $Q$ must hold in a final state,
we establish additional conditions at intermediate steps, with the aim to reason by induction on steps of execution.
The second idea is to designate which intermediate steps in terms of $C$'s CFG.
An \dt{assertion} is thus a formula $R$ associated with a particular point in the CFG, designating the claim that in any run, $R$ holds whenever control is at that point.  This beautiful idea, 
called \dt{annotation}, 
has a simple representation in syntax which has become commonplace: the assert statement.
The third idea ensures that the claim is strong enough to be an induction hypothesis to prove the spec: 
(i)
The entry point is annotated as $P$ and the exit point is annotated as $Q$.
(ii) Every cycle in the CFG is annotated with at least one assertion.  
The annotated points are said to form a \dt{cutpoint set}.  
Such an annotation determines a finite set of acyclic paths through the CFG,
each starting and ending with an annotation and having no intervening one---we call these \dt{segments}.


Floyd shows, by induction on execution steps, that $C$ satisfies $\spec{P}{Q}$ provided that the 
\dt{verification conditions} (VCs) all hold~\cite{Floyd67short}.   
Each segment determines the following VC: for any state that satisfies the initial assertion, and any execution along the segment, such that the branch conditions hold, the final assertion holds in the last state.
In effect, the VCs are cases of the induction step in a proof that the assertions  hold
in any run from a state satisfying $P$.

Given a program and cutpoint set for its CFG, there is an automaton with $Ctrl$ the cutpoint set;
the transitions $(c,s)\trans (d,t)$ are given by the semantics of a segment  from $c$ to $d$.
An annotation assigns a store predicate $anno(c)$ to each cutpoint $c$.
Define the state set $S\subseteq (Ctrl\times Sto)$ by 
\( (c,s)\in S \mbox{ iff } s \models anno(c)
\).
Then the VCs amount to the condition that $S$ is closed under $\trans$.

The IAM requires us to reason about the semantics of straight-line program fragments, which is amenable to automation in a number of different ways, including the direct use of operational semantics~\cite{JMooreVC}.
What makes program verification difficult is finding inductive intermediate assertions.

\subsection{Hoare logic}

Hoare showed that the IAM can be presented as a deductive system, in which inference rules capture the semantics of the constructs of the programming language
and verification conditions are, to some degree, compositonal in terms of program syntax.  
The system derives what are known variously as \emph{partial correctness assertions}, Hoare triples, etc.,
and which ascribe a spec to a program.  
Hoare wrote $P\{C\}Q$ but it has become standard to write $\{P\}C\{Q\}$.
We write $C:\spec{P}{Q}$ and call it a \dt{correctness judgment}.    

There are many Hoare logics, 
because a deductive system is defined for a particular language, i.e., set of program constructs. 
The textbook by Apt et al~\cite{AptOld3} has logics encompassing procedures, concurrency, etc.  
In this paper we focus on sequential programs but the principles apply broadly. 

\begin{figure}[t]
\begin{mathpar}


x:=e : \spec{\subst{P}{x}{e}}{P}

\inferrule
{
C:\spec{P}{R} \\ D:\spec{R}{Q}
}{
C;D : \spec{P}{Q} 
}

\inferrule
{
C:\spec{P\land e}{Q} \\ D:\spec{P\land \neg e}{Q}
}{
\ifc{e}{C}{D} : \spec{P}{Q} 
}

\inferrule
{
C:\spec{P}{Q} \\ D:\spec{P}{R}
}{
C\choice D : \spec{P}{Q\lor R} 
}

\inferrule
{
C:\spec{P\land b}{P} 
}{
\whilec{b}{C} : \spec{P}{P\land \neg b} 
}
\end{mathpar}
\vspace*{-4ex}
\caption{Syntax-directed rules of HL for simple imperative programs.}\label{fig:syntax-directed}
\end{figure}

\begin{figure}[t]
\begin{mathpar}
\inferrule*[right=(Conseq)]{
P\imp R \\ 
C : \spec{R}{S} \\
S\imp Q
}{
C : \spec{P}{Q} 
}

\inferrule*[right=(Conj)]{
C : \spec{P}{Q} \\
C : \spec{P}{R} 
}{
C : \spec{P}{Q\land R} 
}

\inferrule*[right=(Disj)]{
C : \spec{P}{R} \\
C : \spec{Q}{R} 
}{
C : \spec{P\lor Q}{R} 
}

\inferrule*[right=(Frame)]{
C : \spec{P}{Q} \\
FV(R)\intersect Vars(C)=\emptyset
}{
C : \spec{P\land R}{Q\land R} 
}
\end{mathpar}
\vspace*{-4ex}
\caption{Rules to manipulate specs in HL.}\label{fig:structrules}
\end{figure}

Rules for some program constructs can be found in Fig.~\ref{fig:syntax-directed}.
The axiom for assignment involves capture-avoiding substitution of an expression for a variable, written $\subst{P}{x}{e}$, wherein we see that the system treats program expressions and variables as mathematical ones, a slightly delicate topic that does not obtrude in the sequel.  
These rules transparently embody the reasoning that underlies the IAM.
The sequence rule adds an intermediate assertion, changing one verification condition into two (typically simpler) ones. 
The rule for conditional alternatives has two premises, 
corresponding to the two paths through the CFG.
Nondeterministic choice (notation $\choice$) again gives rise to two paths.
The rules in Fig.~\ref{fig:syntax-directed} provide for deductive proof following the program structure
but are incomplete; e.g., 
they give no way to prove the judgment $x:=x+1 : \spec{(x = y)}{(x > y)}$.  
Such gaps are bridged by rules like those in Fig.~\ref{fig:structrules}.
For HL to be a self-contained deductive system it needs to include means to infer valid formulas,
such as the first and third premises of rule \rn{Conseq}.

For while programs, the syntax-directed rules together with \rn{Conseq} are complete in the sense that any true
judgment can be proved.\myendnote{See for example Apt et al~\cite{AptOld3} where 
   completeness is defined relative to completeness of reasoning about assertions and 
  with the requirement that the assertion language be expressive enough to capture loop 
  invariants, which I gloss over in this paper.}
The other rules embody useful reasoning principles.
\rn{Disj} provides proof by cases, which is useful when a program's behavior has two quite different cases.
In terms of IAM, one might apply the method twice, to prove 
$C : \spec{P}{R}$ and $C : \spec{Q}{R}$ with entirely different annotations,
and then conclude $C : \spec{P\lor Q}{R}$ by some argument about the meaning of specs.
This principle is expressed directly in Hoare logic,
as is the oft-used principle of establishing conjuncts of a postcondition separately.

\paragraph{Modular reasoning.}

HL easily admits procedure-modular reasoning, sometimes formalized 
by judgments of the form $H\proves C:\spec{P}{Q}$ where hypothesis $H$
comprises procedure signatures and their specs \cite{OHearnYangReynoldsInfoTOPLAS,RegLogJrnII}.  
With the addition of procedures, \rn{Conseq} is not sufficient 
for completeness. Other rules are needed to manipulate specs,
such as substitution rules to adapt procedure specs to their 
calling contexts~\cite{Hoare71proc,Olderog83adapt,AptOld3}.
We use the name \rn{Frame} for a rule Hoare called Invariance~\cite{Hoare71proc}, with a nod towards a similar rule 
in separation logic~\cite{OHearnReynoldsYangPtrs} where disjointness of heap locations is expressed by using the separating conjunction in place of $\land$.  
With explicit frame conditions the rule can be phrased like this:
From $C : \fspec{P}{Q}{\ol{x}}$ and $FV(R)\intersect \ol{x}=\emptyset$
infer $C: \fspec{P\land R}{Q\land R}{\ol{x}}$.
The principle here is to reason ``locally'' with assertions $P,Q$ pertinent to the effect of 
$C$, and then infer a spec $\spec{P\land R}{Q\land R}$ needed to reason about 
a larger program of which $C$ is a part.
Locality is important for effective reasoning about programs involving the heap.
(Explicit frame conditions for the heap can be found, for example in the Dafny language~\cite{Leino10} and in
the variation of HL dubbed region logic~\cite{RegLogJrnII}.)
The notion of \emph{adaptation completeness} characterizes the extent to which a HL has sufficient rules for reasoning about specs~\cite{Olderog83adapt,KleymannFAC,Naumann01,AptO19}.\myendnote{There seems to 
be only a little room for variation in formulating the syntax-directed rules, 
but more room for the rules that manipulate correctness judgments.
For example, consider this sound rule which is akin to the \rn{Disj} rule:
\[
\inferrule{
C : \spec{P}{Q} \\
x \notin FV(Q)\\
x \notin Vars(C)
}{
C : \spec{\some{x}{P}}{Q}
}
\]
In some sources one finds a variation with conclusion of this form: 
$C : \spec{\some{x}{P}}{\some{x}{Q}}$.  
The variation can be derived from the displayed rule using \rn{Conseq}
and the tautology $Q\imp\some{x}{Q}$.
}

\paragraph{Refinement.}

Validity of assertions is a separate concern from program correctness and \rn{Conseq} 
brings the two together in a simple way---but it has nothing to do with a specific command.
It connects two specs, in a way that can be made precise by defining the \dt{intrinsic refinement order}
($\rt$, ``refined by'') on specs.
Fixing a class of programs, we define
$\spec{P}{Q} \rt \spec{R}{S}$ iff
$C : \spec{R}{S} \mbox{ implies } C : \spec{P}{Q} $ for all $C$.
This relation can itself be given a deductive system, with rules including 
that $\spec{P}{Q} \rt \spec{R}{S}$ can be inferred from 
$P\imp R$ and $S\imp Q$.
The program correctness rule infers
$C: spec_1$ from 
$C : spec_0$ and 
$spec_1 \rt spec_0$.
Using frame conditions, the \rn{Frame} rule can also be phrased as a spec refinement:
\( \fspec{P\land R}{Q\land R}{\ol{x}} \rt \fspec{P}{Q}{\ol{x}} \mbox{ provided }FV(R)\intersect \ol{x}=\emptyset
\).

Disentangling spec reasoning from reasoning about the correctness judgment 
helps clarify that that adaptation completeness is about spec refinement~\cite{Naumann01}.
But it does come at a cost:
To account for the \rn{Conj} and \rn{Disj} rules one needs not only the relation $\rt$ on specs but also
meet/join operators.\myendnote{Such operators are available in JML and are useful in connection with behavioral subtyping~\cite{Leavens-Naumann15}.
One can also define an intrinsic refinement order on programs in terms of the 
   specs they satisfy.
   Tony Hoare long ago shifted his attention towards algebraic approaches in which programs and specs are 
   freely combined and related by refinement.
   In the variation called refinement calculus~\cite{Back:book,Morgan:book,Morris:ref},
   specs are considered as a kind of atomic command the central judgment is refinement, of 
   programs, specs, and combinations thereof,
   which subsumes the correctness judgment $C:\spec{P}{Q}$ as $\spec{P}{Q} \rt C$.  
}
Explicit formalization of spec refinement could be useful for relational specs, 
owing to the additional manipulations that exist owing to the additional dimension.
But I do not develop the topic further in this paper.

\paragraph{Program transformation.}

Verification tools employ semantics-preserving transformations as part of the 
process of generating VCs.\myendnote{For example, rewriting potentially 
faulting or diverging expressions into primitive commands, so other expressions are pure 
and have mathematical semantics as needed by the ``pun'' of program expressions in formulas.
Another example is assigning subexpressions to temporary variables in order for atomic commands to make at most one heap access.  
}
Less commonly, transformations are an ingredient in a Hoare logic.  
An instance of this is the logic for distributed programs of Apt et al~\cite{Apt86,AptOld3},
where the main rule for a distributed program has as premise the correctness of a derived sequential program. 
Another instance is rules to justify the use of auxiliary or \dt{ghost} state in reasoning
\cite{OwickiGries}.
One such rule uses variable blocks $\varblock{\ol{x}}{C}$. 
The variables $\ol{x}$ are called \dt{auxiliary in} $C$ provided their only occurrences 
are in assignments to variables in $\ol{x}$.  
Writing $C \setminus \ol{x}$ for the command obtained by replacing all such assignments with $\skipc$, the rule is 
\[
\inferrule*[right=(AuxVar)]{
\ol{x} \notin FV(P,Q) \\
\ol{x}\mbox{ auxiliary in }C \\
\varblock{\ol{x}}{C} : \spec{P}{Q} 
}{
C\setminus\ol{x} :\spec{P}{Q} 
}
\]
It is sound because the auxiliary variables cannot influence values or branch conditions,
and thus have no effect on the variables in $P$ or $Q$, nor on termination.\myendnote{We give 
the rule \rn{AuxVar} in a simple form, for clarity.
But one may want other ghost computation besides assignments.  For soundness of such a rule the ghost code must 
be terminating~\cite{FilliatreGP16}.}
As relational correctness judgments can express both dependency and program equivalences, we should be able to bring both the condition ``auxiliary in'' and the transformation $C \setminus \ol{x}$ into the logic, making the above rule admissible.


\section{\mbox{Relational properties, alignment, and program products}}\label{sec:align}

Here are some example relational properties of a single program.
\begin{description}
\item[(determinacy)]
For all terminated traces $\tau,\upsilon$ from the same initial state, the final states are the same:
$\tau_0 = \upsilon_0$ implies $\last{\tau} = \last{\upsilon}$.
\item[(monotonicity)] 
For all terminated traces $\tau,\upsilon$, if $\tau_0(x)\leq \upsilon_0(x)$ then $\last{\tau}(z) \leq \last{\upsilon}(z)$.
Here $x,z$ are integer variables.
\item[(dependence, non-interference)] (``$z$ depends on nothing except possibly $x$'')
For all terminated traces $\tau,\upsilon$,
if $\tau_0(x) = \upsilon_0(x)$ then $\last{\tau}(z) = \last{\upsilon}(z)$.
\end{description}
Here are some example relations between programs $C$ and $D$.
\begin{description}
\item[(equivalence)]
For all terminated traces $\tau$ of $C$ and $\upsilon$ of $D$,
if $\tau_0 =  \upsilon_0$ then $\last{\tau} = \last{\upsilon}$.
Determinacy is self-equivalence in this sense.
\item[(majorization)]
For all terminated traces $\tau$ of $C$ and $\upsilon$ of $D$,
if $\tau_0(x) = \upsilon_0(x)$ then $\last{\tau}(z) > \last{\upsilon}(z)$
\item[(refinement)]
For all terminated traces $\tau$ of $C$,
there is a terminated trace $\upsilon$ of $D$
with $\tau_0 =  \upsilon_0$ and $\last{\tau} = \last{\upsilon}$.
\item[(relative termination)]
(For a given relation $\R$.) 
For all initial states $\beta,\gamma$ that satisfy $\R$,
if $C$ has a terminated trace from $\beta$ then 
$D$ has a terminated trace from $\gamma$ ~\cite{hawblitzelklr13}.
\item[(mutual termination)]
For all initial $\beta,\gamma$ that satisfy $\R$,
$C$ can diverge from $\beta$ iff $D$ can diverge from $\gamma$~\cite{GodlinS08}.
\end{description}
Refinement and relative termination involve existential quantification over traces, 
as do generalizations of refinement such as simulation and also 
dependence for nondeterministic programs (if $\tau$ is terminated and $\tau_0(x)=\gamma(x)$ 
then there is terminated $\upsilon$ with $\upsilon_0=\gamma$ and $\last{\upsilon}(z)=\last{\tau}(z)$).
We refer to these as $\forall\exists$ properties, by contrast with the preceding items which 
universally quantify traces (denoted $\forall\forall$).
The $\forall\forall$ properties above are also \dt{termination-insensitive} in the sense that they only constrain terminating traces.  
In this paper we focus on termination-insensitive $\forall\forall$ properties while discussing some $\forall\exists$ properties (which are hyperliveness~\cite{ClarksonSchneiderHyper10}, not 2-safety) in passing.  
Mutual termination also involves existentials, unless programs are deterministic 
as they are in Benton~\cite{Benton:popl04} where mutual termination is used.

Let $A' = (Ctrl',Sto',init',\fin',\trans')$ be an automaton.
A relational spec $\rspec{\R}{\S}$
is comprised of relations $\R$ and $\S$ from $Sto$ to $Sto'$.
We write $(c,s),(c',s')\models \R$ to mean $s,s'\models\R$.
Finite traces $\tau$ of $A$ and $\tau'$ of $A'$ satisfy 
$\rspec{\R}{\S}$,
written $\tau,\tau'\models \rspec{\R}{\S}$,
just if $\tau_0,\tau'_0\models \R$, $ctrl(\last{\tau})= \fin$, and $ctrl(\last{\tau'})=\fin$ imply
$\last{\tau},\last{\tau'}\models \S$.
The non-stuck semantics of relational specs requires, in addition, 
that $ctrl(\last{\tau})\neq fin$ implies $\last{\tau}\trans -$
and  $ctrl(\last{\tau'})\neq fin'$ implies $\last{\tau'}\trans' -$.
Finally, the pair $A,A'$ satisfies $\rspec{\R}{\S}$
just if all pairs of finite traces do, and we write $A|A':\rspec{\R}{\S}$ for satisfaction. 
(Where I write $A|A'$, as in~\cite{MuellerKS15,BanerjeeNN16short}, 
Francez writes $A\times A'$, and  Benton's $A\sim A'$ is popular.)


A key idea (in~\cite{Francez83} and elsewhere) is to form a single automaton, 
runs of which encode pairs of runs of the considered programs, and to which IAM can be applied.
For a single program there is not much flexibility in how it is represented as an automaton or CFG 
but there are many product automata for a given pair of programs---these represent different ways of aligning the steps of the two programs.\myendnote{Francez and others notice the analogy with disjoint 
  parallelism and use terms like ``synchronization''.
  I prefer the term ``alignment''~\cite{KovacsSF13}
  which is apt and not confusing even if relations are considered between concurrent programs.  
}
This flexibility is crucial for the effectiveness of the IAM, specifically on the 
simplicity of annotations and thus the ease of finding them
and proving the VCs.
To discuss this we consider the four examples in Fig.~\ref{fig:examples}.

\begin{figure}[t]
\begin{lstlisting}
P0: (* z := x! *)  y:= x; z:= 1; while y <> 0 do z:= z*y; y:= y-1 od

P1: (* z := \expx *)  y:= x; z:= 1; while y <> 0 do z:= z*2; y:= y-1 od

P2: (* z := x!, half as fast *)
y:= x; z:= 1; w:= 0;
while y <> 0 do if w mod 2 = 0 then z:= z*y; y:= y-1 fi;  w:= w+1 od

P3: (* z := \expx, a third as fast *)
y:= x; z:= 1; w:= 0; 
while y <> 0 do if w mod 3 = 0 then z:= z*2; y:= y-1 fi; w:= w+1 od
\end{lstlisting}
\caption{Example programs. $P0$ and $P1$ are from~\cite{Francez83}.}\label{fig:examples}
\end{figure}

Consider proving monotonicity of $P0$.  To express relations we use dashed ($'$) identifiers for the second run,
so the spec can be written $\rspec{x\leq x'}{z\leq z'}$.
One can prove the functional property that $P0$ computes factorial ($x!$)
and then prove monotonicity for the recursive definition of $!$.
But, as pointed out in~\cite{Francez83}, one can also consider two runs from initial values $x,x'$ with $x\leq x'$,
aligning their iterations in lockstep with invariant $y\leq y'\land z\leq z'$ and no use of $!$.

Consider proving that $P2$ is equivalent to $P0$, which we again specify
just using the relevant variables: $P0|P2: \rspec{x=x'}{z=z'}$.
Lockstep alignment of their iterations is not helpful; we would like to align each iteration of $P0$ 
with two iterations of $P2$ in order to use simple annotations like $y=y'\land z=z'$.

\subsection{Product automata represent alignments}

Let $\otimes$ denote the cartesian product of relations,\myendnote{That is, 
  $(a,b)(R\otimes S)(c,d)$ iff $aRc$ and $bSd$.}
so $\trans\otimes\trans'$ is a relation on $(Ctrl\times Sto)\times(Ctrl'\times Sto')$, i.e., on state pairs.
Let $\id_A$ be the identity relation on states of $A$.
A \dt{pre-product} of $A$ and $A'$ is an automaton $P_{A,A'}$ of the form
\( ((Ctrl\times Ctrl'), (Sto\times Sto'), (init,init'), (\fin,\fin'), \biTrans)
\)
such that 
we have 
${\biTrans} \subseteq (\trans\otimes\trans') \union (\trans\otimes \id_{A'}) \union (\id_{A} \otimes \trans')$.
The union is disjoint, owing to non-stuttering of $A$ and $A'$.
Each transition of $P_{A,A'}$ corresponds to one of both $A$ and $A'$, or else one of $A$ or $A'$ leaving the other side unchanged.  
Such $\biTrans$ satisfies the requirements of finality and non-stuttering.

Let $T$ be a trace of a pre-product of $A,A'$.  Mapping the first projection ($\mathconst{fst}$) over $T$ does not necessarily yield a trace of $A$, as it may include stuttering steps (related by $\id_A$).
So we define $\Left(T)$ to be $\destutter(\mathconst{map}(\mathconst{fst},T))$ where $\destutter$ removes stuttering transitions.\myendnote{For trace $\tau$ define
$\destutter(\tau) = \tau$ if $len(\tau)=1$,
$\destutter(\gamma_0::\gamma_1::\tau) = \destutter(\gamma_1::\tau)$
if $\gamma_0 = \gamma_1$, 
and $\destutter(\gamma_0 :: \gamma_1 :: \tau) = \gamma_0::\destutter(\gamma_1::\tau)$
otherwise.
}
Observe that $\Left(T)$ is a trace of $A$, and we obtain \emph{mutatis mutandis} a trace, $\Right(T)$, of $A'$.
A \dt{pre-product} is \dt{adequate} if it covers all finite traces:
For all finite traces $\tau$ of $A$ and $\tau'$ of $A'$ there is a trace $T$ of $P_{A,A'}$ 
with 
$\tau\prefix\Left(T)$ and $\tau'\prefix\Right(T)$,
where $\prefix$ means prefix.   
It is \dt{weakly adequate} if it covers all finite prefixes $\tau,\tau'$ of terminated traces.
(To see that equality $\tau= \Left(T)$ and $\tau'=\Right(T)$ would be too restrictive,
consider lockstep alignment with $\tau$ strictly shorter or longer than $\tau'$.)

Owing to the definition of states of a pre-product,
a relational spec $\rspec{\R}{\S}$ for $A,A'$ can be seen as a
unary spec $\spec{\R}{\S}$ for $P_{A,A'}$.
For a trace $T$ of $P_{A,A'}$ we have 
$T\models \spec{\R}{\S}$ iff $\Left(T),\Right(T)\models \rspec{\R}{\S}$ by definitions.
We obtain the following by definitions.
\begin{theorem}\label{thm:product}
For the basic semantics of specs,
if $P_{A,A'}$ is a weakly adequate pre-product of $A,A'$
then $P_{A,A'}$ satisfies $\spec{\R}{\S}$ iff 
the pair $A,A'$ satisfies $\rspec{\R}{\S}$.
\end{theorem}
This confirms that a relational spec can be proved using the IAM
and a pre-product.  The challenge is to construct one that admits a simple annotation and is at least weakly adequate.   Adequacy is a more robust condition that holds for several forms of product.  
The number of cutpoints needed for a product may be on the order of the product of the number for 
the underlying automata, but a good alignment makes many unreachable; those can be annotated as false
so the corresponding VCs are vacuous.

Apropos stuck states, an adequate pre-product may have stuck states that do not correspond to stuck states of $A$ or $A'$;
this is one way for a pre-product to be helpful, rendering unreachable states such as those where the guards of a conditional are not in agreement.
However, 
if $P_{A,A'}$ is an adequate pre-product and satisfies $\spec{\R}{\S}$ in non-stuck semantics, 
it does not follow that $A,A'$ satisfies $\rspec{\R}{\S}$ in non-stuck semantics ---contrary to a misstatement in a previous version of this paper.
The problem is one-sided divergence.  For example, suppose trace $\tau$ of $A$ is not terminated but $\last{\tau}$ is stuck.
If $A'$ can diverge then it is possible for a $P_{A,A'}$ to be adequate yet have an infinite sequence of traces $T_i$ such that $i<j \imp T_i < T_j$ (so that it is not stuck), yet $\mathconst{fst}(T_i)=\tau$ for all $i$.
That problem can be solved by requiring $A'$ to be terminating (and $A$ as well, to prevent divergence on the left), an approach taken in~\cite{BanerjeeNN16}.
Alternatively, products can be required not to have one-sided divergence,
an approach taken in~\cite{NikoueiBN19a}.

Here are some pre-products defined for arbitrary $A,A'$.
For brevity, we express product states as pairs of $A$- and $A'$-states,
as if the product's state had type 
$(Ctrl\times Sto)\times (Ctrl'\times Sto')$.
\begin{description}
\item[only-lockstep.] 

$(\gamma,\gamma')\biTrans_{olck} (\beta,\beta')$ iff $\gamma\trans \beta$ and $\gamma'\trans' \beta'$.

\item[eager-lockstep.]
$(\gamma,\gamma')\biTrans_{elck} (\beta,\beta')$ iff
$(\gamma,\gamma')\biTrans_{olck} (\beta,\beta')$, or 
$\ctrl(\gamma)=\fin$ and $\gamma'\trans' \beta'$ and $\gamma=\beta$,
or
$\ctrl(\gamma')=\fin'$ and $\gamma\trans \beta$  and $\gamma'=\beta'$.


\item[interleaved.]
%
$(\gamma,\gamma')\biTrans_{int} (\beta,\beta')$ iff
$\gamma\trans \beta$ and $\gamma'=\beta'$ or
$\gamma'\trans' \beta'$ and $\gamma=\beta$.

\item[maximal.] The union $\biTrans_{olck} \union \biTrans_{int}$.

\item[sequenced.] 
$(\gamma,\gamma')\biTrans_{seq} (\beta,\beta')$ iff 
$\gamma\trans \beta$ and $\ctrl(\gamma')=init'$ and $\gamma' = \beta'$ 
or 
$\ctrl(\gamma)=\fin$ and $\gamma= \beta$ and $\gamma'\trans' \beta'$.

\item[simple-condition.]
%
Given ``alignment condition'' 
$ac \subseteq (Ctrl\mathord{\times} Sto)\times(Ctrl'\mathord{\times} Sto')$,
define $\biTrans_{scnd}$ by 
$(\gamma,\gamma')\biTrans_{scnd} (\beta,\beta')$ iff
either 
$(\gamma,\gamma')\in ac$ and 
$(\gamma,\gamma')\biTrans_{olck} (\beta,\beta')$
or $(\gamma,\gamma')\notin ac$ and 
$(\gamma,\gamma')\biTrans_{int} (\beta,\beta')$.

\end{description}
As Francez observes, interleaved has a relatively large reachable state space, making it more difficult to find inductive invariants.

The only-lockstep form is not adequate, in general, because a terminated state or stuck state can be reached on one side before it is on the other.  
The eager-lockstep, interleaved, and maximal pre-products are all adequate.
The sequenced form is not adequate in general: a stuck or divergent state on the left 
prevents coverage on the right.
Sequenced is weakly adequate if $A,A'$ have no stuck states.

The simple-condition product can also fail to be adequate: if $ac$ holds, both sides are required to 
take a step, which may be impossible if one side is stuck or terminated.
It is also insufficiently general: as we show later, it may be most convenient to designate that steps should be taken on one side or the other.  This suggests the following, which subsumes the preceding constructions.
\begin{description}
\item[3-condition.]
Given state conditions $l,r,b$, define 
$(\gamma,\gamma')\biTrans_{3cnd} (\beta,\beta')$ iff
either 
$(\gamma,\gamma')\in l$ and $\gamma\trans\beta$ and $\gamma'=\beta'$, or 
$(\gamma,\gamma')\in b$ and $(\gamma,\gamma')\biTrans_{olck} (\beta,\beta')$, or 
$(\gamma,\gamma')\in r$ and $\gamma'\trans'\beta'$ and $\gamma=\beta$.
\end{description}

\subsection{Examples}\label{sec:examp}

Consider proving that $P0$ majorizes $P1$, for inputs $x>3$, that is,
\( P0|P1: \rspec{ x=x'\land x > 3 }{z > z'} \).\myendnote{In~\cite{Francez83}
this is claimed for $x>2$, but that is not true of these programs, as we have $3! < 2^3$.}  
Francez observes that using sequenced product would require reasoning about $z=x!$ and $z'=2^{x'}$,
and suggests aligning the iterations in lockstep and using this relational invariant:
\( y=y' \land (z=z'=1 \lor z > z') 
\).
This condition is not preserved by the loop bodies under guard condition $y>0$,
for example in the state $y=2,z=6,z'=4$ reached when $x=x'=3$, but here we are concerned with the case $x>3$.
If we add $x>3$ as a conjunct we get a condition that is indeed invariant for lockstep executions,
but it is not inductive--- that is, the verification condition for the loop body is not valid.  
But there is a simple invariant with which the relation can be proved:
\begin{equation}\label{eq:majinv}
y=y' \land ((y > 4 \land z = z' = 1) \lor (y > 0 \land z > 2*z') \lor (y = 0 \land z > z')) 
\end{equation}
This is not established by the initialization, in case $x = 4$. 
Instead we use this invariant to prove correctness under precondition $x>4$ 
and separately prove correctness under the very specific precondition $x=4$ which can be proved, for example, by unrolling the loops.  In short, we do case analysis, as in rule \rn{Disj}.\myendnote{In the case 
  we have precondition $x>4$, the initializations establish (\ref{eq:majinv}), in particular the first of the disjuncts.
  Note that following the loop we have both (\ref{eq:majinv}) and the negated guard, i.e. $y=0$, 
  whence the postcondition $z>z'$.
  It remains to show that (\ref{eq:majinv}) is preserved by the loop body
  when $y\neq 0$.
  To this end we reason forward, 
  considering each disjunct in turn, and writing $y_0,z_0,y'_0,z'_0$ for the initial values.
  so we have $y=y_0-1$, $y'=y_0'-1$, $z=z_0*y_0$, and $z'=z_0*2$.
  \begin{itemize}
  \item If the first disjunct holds initially, i.e., $y_0>4 \land z_0=z'_0=1$,
  we get $z = y_0 > 4 = 2*2 = z'*2$ and $y>0$, whence the second disjunct in (\ref{eq:majinv}).
  \item If the second disjunct holds initially, i.e., $y_0>0 \land z_0 > 2*z'_0$, 
  we make a further case split:
     \begin{itemize}
     \item if $y_0 = 1 \land z_0 > 2*z'_0$ 
         then $y=0$ and we get the rest of the third disjunct by $z = z_0*1 > 2*z'_0 = z'$.
     \item if $y_0 > 1 \land z_0 > 2*z'_0$ then $y>0$ and we get the rest of the second disjunct by
        $z = z_0*y_0 \geq z_0*2 > 2*z'_0*2 = 2*z'$
     \end{itemize}
  \item The third disjunct does not hold initially, given the guard conditions $y\neq 0$.
  \end{itemize}
} 

Program $P2$ is equivalent to $P0$, and $P3$ to $P1$, but neither fact is easily proved 
using lockstep alignment.  
For the simplest invariants in proving $P0$ equivalent to $P2$ we should let $P2$ take two iterations for each one of $P0$.
The question is how to formulate that nicely.

As another example, $P2$ majorizes $P3$, for $x>4$,
but again this is not easily proved by reasoning about lockstep alignment of the loops.  
Both programs have gratuitous iterations in which $y$ and $z$ are not changed.
We would like to align the computations so that when $w=w'=0$ 
we can assert (\ref{eq:majinv}).
Indeed, when $w\neq 0$ (respectively $w'\neq 0$), an iteration on the left (resp.\ right)
has no effect on the other variables and thus maintains (\ref{eq:majinv}).
For this proof we may try a simple-condition product 
so joint steps are taken when \lstinline{(w mod 2) = 0 = (w' mod 3)}.
But this is insufficient: it allows one side to run ahead 
in states where the condition does not require both sides to step together,
precluding a simple invariant.
What we need is a 3-condition product.
The left may advance independently when \lstinline{w mod 2 <> 0} and \lstinline{w/2=w'/3};
the right when \lstinline{w' mod 3 <> 0} and \lstinline{w/2=w'/3}. 
Then (\ref{eq:majinv}) is invariant.

The examples only scratch the surface. Compilation, refactoring, and program revision need less obvious alignments, but often do admit alignments for which simple and even inferrable invariants suffice.

In examples like equivalence of $P0$ and $P2$  there is a fixed correspondence between loops of one versus the other program, a pattern that arises in some loop transformations used in compilers (e.g., to introduce vector operations).  
For majorization of $P3$ by $P2$ our alignment is more data-dependent, although it is not surprising 
that it can be described succinctly since the iterations have a regular pattern.
Here is a less regular example (from~\cite{BanerjeeNN16short}): 
the program uses a loop to sum the integers in a list, where list elements have a boolean flag that indicates an element should be ignored.  The property is that two runs yield the same sum, provided the two lists have the same non-deleted elements in the same order.  
This can be handled nicely using a 3-condition product.\myendnote{Advance on the left (resp.\ right) if the next element on the left (resp.\ right) is deleted; if neither are deleted then both sides advance together.}

One can imagine more elaborate product automata using ghost state to track alignment conditions,
but it seems that in any case what is needed is to designate when to advance on one side, the other side, or both. 

\section{Rules of relational program logic}\label{sec:RHL}

As has been rediscovered and noted several times, it is not difficult to 
use program syntax to make a program that behaves as a product of programs.
A simple case, mentioned earlier, is the sequence $C;C'$ where $C'$ has no variables in common with $C$, and which corresponds to the sequenced product automaton.
But it is also natural to interleave code from such disjoint programs, so as to align 
intermediate points in control flow.  
For a deductive system one also needs to account for the connection between such a product and the original program (or programs), the primary objects of interest.
It is also desirable to disentangle reasoning principles, such as various alignments, from details of encoding.
Furthermore, although disjoint variables suffice to reduce relational reasoning to standard HL for simple imperative programs, this is no longer the case for languages with more elaborate notions of state.
For example, many languages feature a single heap and it is not trivial to use it to encode two disjoint heaps (see~\cite{Naumann06esorics,Beringer11}).  
Another example is assembly language for a conventional architecture with a fixed set of registers.
In such situations it may be preferable to work more directly with the relational correctness judgment,
suitably interpreted, rather than depending entirely on products encoded as single programs.

We have reached the main topic of this paper, deductive systems for the relational judgment $C|C':\rspec{\R}{\S}$,
in which various principles of reasoning are manifest in proof rules.
With HL in mind we may expect 
syntax-directed rules that embody program semantics,
rules for manipulation of specs, 
and rules for program transformation.
In addition, relational reasoning benefits from judicious alignment of program fragments.
For lockstep automata, the corresponding rules are dubbed ``diagonal''~\cite{Francez83}
and relate programs with the same control structure.
The sequenced and interleaved automata involve one-sided steps, corresponding 
to proof rules syntax-directed on one side.
The 3-condition product is manifest in a three-premise rule for relating two loops.
There are also rules that involve both relational and unary judgments.

Good alignment not only enables use of simple assertions,
it is also essential to enable the use of relational specs for procedure calls.
For lack of space we do not delve into this topic.

We refrain from formalizing relational formulas but we do assume they are closed under the propositional connectives with classical semantics.  
Usual formulations of HL rely on the use of program variables and expressions both as part of programs and as terms in formulas; in relational formulas we need to designate whether they refer to the left or right
execution.
As an alternative to the dashed/undashed convention used in Sec.~\ref{sec:align},
we use the notation $\leftex{e}$ (resp.\ $\rightex{e}$) for the value of expression $e$ on the left (resp.\ right) side.
As naming convention we tend to use dashed names for commands on the right side, but this does
not imply renaming of variables or anything of the sort.
In the logic, the programs are considered to act on distinct states which may or may not have the same variables.
For example, we can write $\leftex{x}\leq\rightex{x}$ rather than $x\leq x'$.  

\subsection{Diagonal and one-side rules}\label{sec:diagonal}

\begin{figure}[t]
\begin{mathpar}

x:=e \sep x':=e' : \rspec{\subst{\R}{x|x'}{e|e'}}{\R}

\inferrule{
  C|C' : \rspec{\R}{\Q} \\
  D|D' : \rspec{\Q}{\S}
}{
  C;D \Sep C';D' : \rspec{\R}{\S}
}

\inferrule{
  C|C' : \rspec{\R\land \leftF{e}\land\rightF{e'}}{\S} \\
  D|D' : \rspec{\R\land \leftF{\neg e}\land\rightF{\neg e'}}{\S} \\
  C|D' : \rspec{\R\land \leftF{e}\land\rightF{\neg e'}}{\S} \\
  D|C' : \rspec{\R\land \leftF{\neg e}\land\rightF{e'}}{\S} 
}{
  \ifc{e}{C}{D} \Sep \ifc{e'}{C'}{D'} : \rspec{\R}{\S}
}

\inferrule*[right=(AltAgree)]{
  \R \imp \leftex{e} = \rightex{e'} \\
  C|C' : \rspec{\R\land \leftF{e}\land\rightF{e'}}{\S} \\
  D|D' : \rspec{\R\land \leftF{\neg e}\land\rightF{\neg e'}}{\S} 
}{
  \ifc{e}{C}{D} \Sep \ifc{e'}{C'}{D'} : \rspec{\R}{\S}
}


\inferrule*[right=(IterAgree)]{
  \Q \imp \leftex{e} = \rightex{e'} \\
  C\sep C' : \rspec{\Q\land \leftF{e}\land\rightF{e'}}{\Q} 
}{ 
  \whilec{e}{C} \Sep \whilec{e'}{C'} : \rspec{\Q}{\Q\land \leftF{\neg e}\land\rightF{\neg e}}
}

\inferrule{
  \Q \imp \leftF{e}=\rightF{e'} \lor (\Lrel\land\leftF{e}) \lor (\R\land\rightF{e'})  \\
  C\sep C' : \rspec{\Q\land \leftF{e}\land\rightF{e'}\land\neg\Lrel\land\neg\R}{\Q} \\
  C\sep\skipc : \rspec{\Q\land\Lrel\land \leftF{e}}{\Q} \\
  \skipc\sep C' : \rspec{\Q\land\R\land \rightF{e}}{\Q} 
}{ 
  \whilec{e}{C} \Sep \whilec{e'}{C'} : \rspec{\Q}{\Q\land \leftF{\neg e}\land\rightF{\neg e}}
}

\end{mathpar}
\vspace*{-4ex}
\caption{Diagonal syntax-directed rules.}\label{fig:diag}
\end{figure}
The rules in Fig.~\ref{fig:diag} relate programs with the same control structure.
Such rules are found in~\cite{Francez83,Benton:popl04,Yang:tcs04} and many other papers.
In the assignment rule, the notation $\subst{\R}{x|x'}{e|e'}$ is meant to be the formula $\R$ in which left-side occurrences of $x$ are replaced by $e$ and right-side occurrences of $x'$ by $e'$.
For example, $\subst{(\leftex{x}=\rightex{x})}{x|x}{x+1|y}$ is $\leftex{x+1}=\rightex{y}$.
The first rule for if/else is general, covering the possible control flows, 
whereas \rn{AltAgree} is applicable when the guard conditions are in agreement 
(and can be understood in terms of simple-condition pre-product with a condition to ensure adequacy).
\rn{AltAgree} can be derived from the first rule, 
using that $C|C':\rspec{\mbox{false}}{\S}$ 
and \rn{RelConseq} (Fig.~\ref{fig:rel-structrules}).

The \rn{IterAgree} rule (e.g., \cite{Benton:popl04,Yang:tcs04}) is applicable when the loop conditions remain in agreement under lockstep alignment; it uses a single invariant relation $\Q$ much like the unary loop rule.
The rule can be use to prove example $P0$ majorizes $P1$, for $x>4$, 
using (\ref{eq:majinv}) as invariant. 
Francez gives a loop rule that corresponds to the eager-lockstep product:\myendnote{As does the rule Fusion 2 of Sousa and Dillig~\cite{SousaD16}.}
with a single invariant like in \rn{IterAgree} but 
with additional premises 
$C\sep \skipc : \rspec{\Q\land \leftF{e}\land\rightF{\neg e'}}{\Q} $
and 
$\skipc\sep C' : \rspec{\Q\land \leftF{\neg e}\land\rightF{e'}}{\Q} $
to handle the situation that one loop continues while the other has terminated;
it is seldom helpful. 
Our second loop rule, from Beringer~\cite{Beringer11}, 
corresponds to the 3-condition product: It augments the invariant $\Q$ 
with two other relations: $\Lrel$ is precondition for an iteration on the left while the right side remains stationary; \emph{mutatis mutandis} for $\R$.  
The side condition $\Q \imp ((\leftF{e}=\rightF{e'}) \lor (\Lrel\land\leftF{e}) \lor (\R\land\rightF{e'}))$ ensures 
adequacy, i.e., covering all pairs of unary traces.


\begin{figure}[t]
\begin{mathpar}

x:=e \sep \skipc  : \rspec{\R^{x|}_{e|}}{\R}


\inferrule*[right=(LeftSeq)]{
  C\sep\skipc : \rspec{\R}{\Q} \\
  D\sep D' : \rspec{\Q}{\S} 
}{
  C;D \sep D' : \rspec{\R}{\S}
}

\inferrule{ 
  B\sep C : \rspec{\R\land \leftF{e}}{\S} \\
  D\sep C : \rspec{\R\land \leftF{\neg e}}{\S} 
}{
  \ifc{e}{B}{D} \Sep C  : \rspec{\R}{\S}
}

\inferrule*[right=(WhSeq)]{
\whilec{e\land b}{B} \sep C : \rspec{\P}{\Q} \\
\whilec{e}{B} \sep D : \rspec{\Q}{\R} \\
\Q\land\leftF{\neg e}\imp\R
}{
\whilec{e}{B} \Sep C;D : \rspec{\P}{\R}
}
\end{mathpar}
\vspace*{-4ex}
\caption{Some left side and mixed structure rules.}
\label{fig:one-side}
\end{figure}

To relate differing programs, a natural idea is one-side rules, some of which we give in 
Fig.~\ref{fig:one-side}.  
The assignment rule is from Francez, where several one-side rules are given with $\skipc$ on the other side, corresponding to interleaved product.
The alternation rule is given in the more general form found in Barthe et al~\cite{BartheCK-FM11,BartheCK16}
and in Beringer~\cite{Beringer11} which also gives \rn{LeftSeq}.
If we identify $D'$ with $\skipc;D'$ (see Sec.~\ref{sec:rel-transform}), rule \rn{LeftSeq}
can be derived from sequence rule in Fig.~\ref{fig:diag}
by replacing $C;D \sep D'$ with $C;D \sep \skipc;D'$.
Right-side rules can be derived using rule \rn{Swap} (Sec.~\ref{sec:rel-specs}).

In addition to one-side rules that relate a structured program with an arbitary one,
Francez considers rules for relating different program structures,
for example \rn{WhSeq}.  
The rule is unusual in that the premises are not judgments for subprograms 
of the one in the conclusion. 
The rule is derivable provided there are rules to rewrite programs to equivalent ones (see Sec.~\ref{sec:rel-transform}).
Since $\whilec{e}{B}$ is unconditionally equivalent to 
the sequence $(\whilec{e\land b}{B}); \whilec{e}{B}$,
rewriting the conclusion results in a relation between two sequences.

\subsection{From unary correctness to relational}\label{sec:unary-to-rel}

If the variables of $C'$ are disjoint from those of $C$ then the semantics of command $C;C'$ amounts to the
sequenced product of the corresponding automata, suggesting:  
\[ 
\inferrule*[right=(SeqProd)]
{ \mbox{$C$ and $C'$ have disjoint variables} \\
            C;C' : \spec{\R}{\S}
}{
C|C':\rspec{\R}{\S}
}
\]
For programs that cannot get stuck, it is sound in basic semantics according to Theorem~\ref{thm:product} and the weak adequacy of sequenced product.
Stuckness can be addressed using additional unary premises.

\rn{SeqProd} is useful as means to obtain relational judgments
for small subprograms such as assignments and basic blocks where a functional spec is not 
difficult to prove.  An alternative way to get relational correctness from unary is 
by this rule, essentially introduced by Yang~\cite{Yang:tcs04}.
\[ 
\inferrule*[right=(Embed)]{
C:\spec{P}{Q} \\
D:\spec{R}{S}
}{
C|D : \rspec{\leftF{P}\land\rightF{R}}{\leftF{Q}\land\rightF{S}}
}
\]
It is sound in both basic and non-stuck semantics.

Typically, the relational assertion language does not express equality of entire states,
but rather of specific variables and sometimes of partial heaps~\cite{Yang:tcs04,BanerjeeNN16short}.
Equivalence of two programs can be specified as $C|C':\rspec{\E}{\F}$ where 
$\E$ (resp.\ $\F$) expresses agreement on whatever parts of the state are 
read (resp.\ written) by $C$ or $C'$.
In a unary logic with frame conditions, suitable $\E,\F$ can be derived from the frame condition~\cite{NikoueiBN19a} but I leave this informal in the following 
rule which yields a relational judgment from a unary one.
\[
\inferrule*[right=(Erefl)]{
C:\spec{P}{Q}
}{
C|C:\rspec{\E\land\Both{P}}{\F}
}
\]
Here $\Both{P}$ abbreviates $\leftF{P}\land\rightF{P}$.
One can add postcondition $\Both{Q}$ by means of \rn{Embed} and \rn{RelConseq}.
Further agreements can be added using \rn{RelFrame} (Fig.~\ref{fig:rel-structrules}).

\subsection{From relational correctness to unary}\label{sec:rel-to-unary}

Preservation of unary correctness by equivalence transformation can be expressed as follows,
where $\E,\F$ are suitable agreements as in \rn{Erefl}.
\[
\inferrule*[right=(Ecorr)]{
C:\spec{P}{Q} \\
C|D : \rspec{\E\land\Both{P}}{\F}
}{
D:\spec{P}{Q} \\
}
\]
Whereas using unary judgments to infer relational ones allows for
a deductive system in which the unary judgment stands on its own,
this rule makes a dependency in reverse.   
We now take a further step which entangles assertion reasoning with correctness judgments.

Francez~\cite{Francez83} motivates interest in the property of monotonicity by considering that it could 
be a requirement on a procedure passed to a routine for numeric integration.
Similarly, a sorting routine requires that the comparator passed to it computes 
a transitive relation, and collections libraries require that the \lstinline{equals} method compute a symmetric relation (at least)~\cite{SousaD16}.
Evidently the functional correctness of such routines relies on
these $k$-safety properties,
but the cited papers do not even sketch such reasoning.
Let us do so, glossing over details about parameter passing.

Consider a sorting routine that uses comparator $comp$ with inputs $x,y$ and output $z$.
Suppose in the proof of $sort(a,comp):\spec{\mbox{true}}{sorted(a)}$ we rely on symmetry.\myendnote{Transitivity
is certain to be needed, but it is 3-safety which is inconvenient for expository purposes.}
That is, some use of \rn{Conseq} is for an entailment that is valid owing to symmetry of comparison.  
Symmetry can be expressed as the relational judgment
$comp|comp : \rspec{x=y'\land y=x'}{z=z'}$.
But we need to connect this with reasoning about unary assertions,
within the confines of a logic of relational and unary correctness judgments.

Such a connection is made in tools and theories that allow ``pure methods'' to be used in 
assertions while avoiding illogical nonsense using arbitrary program functions as mathematical 
ones~\cite{Cok05,BanerjeeNN18}.
Let $C$ be some command meant to compute a function
of input variables $\ol{x}$ as output $z$.
Let $f$ be an uninterpreted (and fresh) name which we will use to represent that function.
We have already seen how to express that $z$ depends only on $\ol{x}$, deterministically:
$C|C: \rspec{\ol{x}=\ol{x}'}{z=z'}$.
A property such as symmetry or monotonicity has the form 
$C|C: \rspec{\R(\ol{x},\ol{x}')}{\S(z,z')}$.
To express that $f$ is the function computed in $z$ we use 
a unary spec, thus $C: \spec{\mbox{true}}{z=f(\ol{x})}$.
Finally, we express the relational property of $f$ as 
a first order (unary) formula:
$\all{\ol{x},\ol{x}'}{ \R(\ol{x},\ol{x}') \imp \S(f(\ol{x}), f(\ol{x}'))}$.
With these ingredients we can state a rule.
\[
\inferrule*[right=(CmdFun)]{
  C|C: \rspec{\ol{x}=\ol{x}'}{z=z'} \\
  C|C: \rspec{\R(\ol{x},\ol{x}')}{\S(z,z')} \\ \mbox{$f$ fresh} \\
  C: \spec{\mbox{true}}{z=f(\ol{x})} \: ; \; 
    (\all{\ol{x},\ol{x}'}{ \R(\ol{x},\ol{x}') \imp \S(f(\ol{x}), f(\ol{x}'))})
    \proves D: \spec{P}{Q}
}{
  \mbox{``link $D$ with $C$''} : \spec{P}{Q}
}
\]
We are glossing over procedures and parameter passing, and termination of $C$.
The last premise, for $D$, is meant to indicate reasoning under a hypothesis.
The hypothesis includes a unary judgment, as in formalizations of HL with procedures.
It also includes the axiom about $f$ for reasoning about assertions.
The rule does not require $C$ to be entirely deterministic and have no effects on other variables besides $z$, but we should disallow writes to $\ol{x}$, so $z=f(\ol{x})$ means what we want.  

From $C: \spec{\mbox{true}}{z=f(\ol{x})}$ 
one can derive $C|C: \rspec{\ol{x}=\ol{x}'}{z=z'}$ by \rn{Embed} and \rn{RelConseq}.
But \rn{CmdFun} does not require proof of $C: \spec{\mbox{true}}{z=f(\ol{x})}$. 
Instead, that spec is used to define $f$ in terms of $C$, in reasoning about $D$.


\subsection{Reasoning about specs}\label{sec:rel-specs}

\begin{figure}[t]
\begin{mathpar}

\inferrule*[right=(RelConseq)]{
  \P\imp \R \\
  C|D : \rspec{\R}{\S} \\
  \S \imp \Q 
}{
  C|D : \rspec{\P}{\Q} \\
}

\inferrule*[right=(RelFrame)]{
C|C' : \rspec{\P}{\Q} \\
FV(\R)\mbox{ disjoint from }Vars(C,C')
}{
C|C' : \rspec{\P\land \R}{\Q\land \R} 
}

\inferrule*[right=(Swap)]{
C|C' : \rspec{\P}{\Q} 
}{
C'|C : \rspec{\P\converse}{\Q\converse} 
}

\inferrule*[right=(Comp)]{
C_0|C_1: \rspec{\P}{\Q} \\
C_1|C_2 : \rspec{\R}{\S}
}{
C_0|C_2: \rspec{\P;\R}{\Q;\S} 
}

\end{mathpar}
\vspace*{-4ex}
\caption{Some rules that manipulate specs.}\label{fig:rel-structrules}
\end{figure}


The reasoning embodied by \rn{Conseq} and other spec rules in HL is also needed in RHL,
e.g., in Sec.~\ref{sec:examp}
we suggested an appeal to the relational disjunction rule.
Some of these rules are in Fig.~\ref{fig:rel-structrules}.
In addition to logical connectives, it is natural to consider formulas with 
converse and relational composition, for which I write $\R\converse$ and $\R;\S$ respectively.  
Rule \rn{Swap} is sound in basic and non-stuck semantics (but not for relative termination).  
Rule \rn{Comp} is not sound in basic or non-stuck semantics, owing to possible divergences of $C_1$;
these are precluded under relative termination and mutual termination semantics.
Soundness of \rn{Comp} can also be achieved using an additional premise for termination.

Let us abbreviate the agreement $\leftex{x}=\rightex{x}$ by $\Agr x$.  
We have focused on local agreements like $\Agr x$, but one may wish to include a global identity relation,
for which we write $\I$.
As Benton shows, partial equivalences (symmetric and transitive relations, \dt{per} for short)
are particularly important, and relation operations let us express such properties as 
valid implications:
$\R\converse\imp\R$ (symmetry) and $\R;\R\imp\R$ (transitivity).
Several works use relational specs to express partial declassification of secrets (e.g.,~\cite{NanevskiBG11short}).
To declassify the value of expression $e$, a typical precondition has the form $\Agr e\land\Both{P}$
which is a per but not reflexive. 
Apropos rule \rn{Comp} instantiated in the form $C_0|C_2: \rspec{\R;\R}{\S;\S}$, 
if $\S$ is transitive we obtain 
$C_0|C_2: \rspec{\R;\R}{\S}$ using \rn{RelConseq}.
Then if $\R$ is reflexive ($\I\imp\R$) 
we obtain $C_0|C_2: \rspec{\R}{\S}$, as $\I;\R$ is equivalent to $\R$.

By analogy with rule \rn{Ecorr} we would like to reason about preservation of a relational property by equivalence transformation.
Consider the relation $C|C' : \rspec{\R}{S}$
together with equivalences $D|C : \rspec{\E}{\F}$ and $C'|D' : \rspec{\E}{\F}$
where $\E,\F$ are suitable agreements.
By \rn{Comp} we get $D|D' : \rspec{\E;\R;\E}{\F;\S;\F}$.  
If $\E$ is a conjunction of agreements including variables of $\R$,
then $\R$ is equivalent to $\E;\R;\E$ and likewise for $\S$ so by \rn{RelConseq} we obtain
$D|D' : \rspec{\R}{\S}$.  
Besides enabling derivation of right-side rules from left-side rules,
rule \rn{Swap} facilitates instantiating the preceding reasoning 
in case $C=C'$ and $D=D'$, to show a security property of $C$ is preserved by the equivalence.
(Take $\R,\S$ to be agreement on non-secret variables.)

Benton~\cite{Benton:popl04} makes the beautiful observation that just as the relational 
spec $\rspec{\Agr x}{\Agr z}$ characterizes a dependency property of a single program,
it also captures that two programs are equivalent with respect to their effect on $z$, 
e.g. $z:=x;y:=z \sep z:=x : \rspec{\Agr x}{\Agr z}$ captures a dead-code elimination transform, 
for a context where the subsequent code does not use $y$ and therefore requires no agreement on it.  

With this in mind, consider programs in which atomic actions happen in different orders,
for example $z:=x+1;w:=y$ versus $w:=y;z:=w+1$, the equivalence of which can be
expressed by the spec $\rspec{\Agr x\land \Agr y}{\Agr z\land \Agr w}$.
A general rule for commuting assignments can be formulated 
requiring disjointness of the variables read in the assignments.
Moreover, one can express such a rule for assignments involving heap locations, 
given means to express agreements thereof.

Heap agreements are often needed up to bijective renaming of 
pointers~\cite{BanerjeeNaumann02cShort,Beringer11}, which can be encoded in ghost state.
%
Such specs can be localized to the locations read and written by a given command, since preservation of additional agreements can be derived by \rn{RelFrame}.
Yang's logic~\cite{Yang:tcs04} features a frame rule taking advantage of separating conjunction of relations.
It is also possible to formulate a frame rule based on relational specs with frame conditions,
as in the work of Banerjee et al~\cite{NikoueiBN19a} which features local equivalence specs derived from frame conditions.



\subsection{Transformations}\label{sec:rel-transform}

The diagonal and one-side rules enable reasoning in terms of convenient alignments but 
apply only to specific control structure patterns.  Programs that do not exactly match the patterns can be rewritten by equivalences such as 
$\skipc; C \uequiv C$, $C; \skipc \uequiv C$, and the following:
\begin{lstlisting}
while e do C od \uequiv while e do C; while e /\ e0 do C od od
while e do C od \uequiv if e do C fi; while e do C od
\end{lstlisting}
Commands $C,C'$ are \dt{unconditionally equivalent}, written
$C \uequiv C'$, if they have exactly the same store traces.\myendnote{That is, projecting out just the stores from their state traces.}
The relation can be formalized using laws like these together with congruence rules.
Such equivalences can be used to desugar fancy control structures,
as done in some verification tools; the justification is that 
$C:\spec{P}{Q}$ and $C\uequiv D$ implies $D:\spec{P}{Q}$ 
(cf.\ rule \rn{Ecorr} in Sec.~\ref{sec:rel-specs}).
The relational logic of Banerjee et al~\cite{BanerjeeNN16short} features a rule like this:
from 
$C|C' : \rspec{\R}{\S}$, $D\uequiv C$, and $C' \uequiv D'$, infer $D|D' : \rspec{\R}{\S}$.
The rule is applied in proving a loop tiling transformation,
using the above rewrites to enable application of diagonal rules.
Transformations are used similarly in~\cite{BartheGHS17,KieferKU18}.
To enable use of sequenced product one may use the equivalence
$\varblock{x}{C} \uequiv \varblock{x'}{\subst{C}{x}{x'}}$ for fresh $x'$.


It seems unparsimonious to rely on an additional program relation ($\uequiv$) for which axioms and rules must be provided and proved sound, in a setting where we already consider a form of program 
relation.\myendnote{Indeed, Benton formulates some unconditional equivalences within his relational calculus DDCC~\cite[sec 3]{Benton:popl04}.}
On the other hand, we have seen in Sec.~\ref{sec:rel-specs} 
that there are limitations on the use of equivalence judgments for reasons of termination.
Having a separate judgment of unconditional equivalence is one way to address termination 
in connection with the basic or non-stuck semantics of relational judgments.

\subsection{Alignment completeness}\label{sec:align-complete}

The usual notion of completeness is that true judgments are provable. 
Suppose the relational judgment $C|C': \rspec{\R}{\S}$ is true.
In a setting where $\R,\S$ can be expressed as, or considered to be, unary formulas,
one can prove it by application of \rn{SeqProd}.
In turn, the sequence can be reduced to true judgments $C:\spec{\R}{\Q}$
and $C':\spec{\Q}{\S}$.  What matters is not that an explicit product $C;C'$ can be 
formed but rather that store relations can be expressed as store  
predicates~\cite{Francez83,BartheDArgenioRezk,BeringerHofmann07,Beringer11}.
If so, the judgment is provable provided the unary HL is complete.
Then a single rule for relational judgments (\rn{SeqProd} or \rn{Embed}) is complete on its own!\myendnote{
The fact that the technique of \rn{SeqProd} is complete relative to unary HL 
   is observed by Francez~\cite{Francez83} but not worked out in detail.
   The fact is also mentioned in Barthe et al~\cite{BartheDArgenioRezk}, for the special case of relating
   a program to itself; it is evident that it holds more generally as noted in~\cite{BartheCK-FM11}.
   Semantic completeness of the technique, for general relational properties, is proved by Beringer~\cite{Beringer11}.
   A RHL is proved complete for deterministic programs, on this basis, in~\cite{BartheGHS17}.
   The logic of Sousa and Dillig~\cite{SousaD16} includes a rule like \rn{SeqProd} for $k$-products,
   and their Theorem 2 is completeness relative to completeness of an underlying HL;
   Wang et al~\cite{WangDilligLahiriCook} prove a similar result specialized to program equivalence.
   The crux of these completeness results is that product programs are complete in the 
sense of representing all pairs or $k$-tuples of unary executions (called adequacy in this paper).  Francez gives a semantic 
completeness result of this sort, for eager-lockstep product, as do Eilers et al~\cite{EilersMH20} 
for a more general form of product.
   For (higher order) functional programs, Aguirre et al~\cite{AguirreBGGS19} prove completeness for a
   RHL via embedding in a unary logic.
} 
A different notion is needed. 

Suppose $C:\spec{P}{Q}$ can be proved using IAM with a particular annotation.
Then there is a HL proof using that annotation, in the sense that at least
the loop rule is instantiated according to the annotation (assuming that loops are cut at loop headers).
Why? Because the VCs will be provable, by completeness of HL,
and the syntax-directed rules suffice to compose the VCs.  
In this sense, HL is complete with respect to IAM for unary correctness.

A natural measure of completeness for RHL is whether 
any proof of $C|C': \rspec{\R}{\S}$ using IAM with a product automaton
can be represented by an RHL proof \emph{using the same annotation and alignment.}
Turning this into a precise definition requires, first, a convincing 
general definition of product automaton; our 3-condition form is relatively general 
but does not encompass the use of ghost state for alignment conditions or store 
relations.
Second, the correspondence between proof rules and aligned products,
discussed informally throughout Secs.~\ref{sec:diagonal}--\ref{sec:rel-transform},
needs to be made precise.
To this end it may help to limit attention to annotations in which all branch points are cutpoints.
We leave this to future work but note that formal proof outlines~\cite{AptOld3}
may be a convenient intermediary.

It is straightforward to add ghost state to our notions of pre-product and adequacy,
to express store relations and alignments.
But some program transformations used in optimizing compilers reorder an unbounded number of atomic actions.
These do not have an obvious representation by pre-product
and they have not been formalized using RHL rules~\cite{NamjoshiS16}. 

\section{Selected additional related work}\label{sec:related}

The idea of relating $C$ to $C'$ by unary reasoning about a program that 
represents their product goes back at least to the 1970s.  
In Reynolds' book~\cite{ReynoldsCraft} we find stepwise refinement from an algorithm 
$C$ using local variables of abstract mathematical types to $C'$ acting on concrete data structures, expressed by augmenting $C$ with parts of $C'$ interwoven in such a way that assertions can express the coupling relation between abstract and concrete data.
DeRoever and Engelhardt call this Reynolds' method and devote a chapter to it~\cite{deRdataref},
citing work by Susan Gerhart~\cite{Gerhart78} as precursor.  
Morgan~\cite{MorganAux} formalizes the idea in terms of auxiliary variables, cf.\ rule \rn{AuxVar}.
The idea of encoding two runs as a sequence of disjoint copies, and specifying determinacy as a Hoare triple, appears (in passing) in a 1986 paper by Csirmaz and Hart~\cite{CsirmazH86}.

Early work by Rinard and Marinov~\cite{RinardMarinov99,RinardCredible99} gives a logical formulation 
of verification conditions for a $\forall\exists$ relation, to prove correctness of compiler transformations
acting on control flow graphs.

The influential papers by Benton~\cite{Benton:popl04} and Barthe et al~\cite{BartheDArgenioRezk}
have been followed by many works.  The following paragraphs give selected highlights.

Barthe, Crespo and Kunz~\cite{BartheCK16} give several ways of formulating 
deductive reasoning about relational properties, 
including deductive systems for product programs in conjunction with unary HL.
They formalize a judgment that connects two commands with a command that represents 
their product.  Products include assertions which must be verified to 
ensure what we call adequacy.


Beringer~\cite{Beringer11} considers partial correctness specs in ``VDM style'' i.e., as relations from pre- to post-state, so partial correctness means the relational semantics of the program is a subset of the spec.
He defines relational decompositions, essentially the relations that hold at the semicolon of a
product $C;C'$ (as in rule \rn{SeqProd}), and observes that given such an ``interpolant'' one can derive VCs for $C$ and $C'$ 
as quotients in the sense of relational calculus (also known as weakest prespecification~\cite{HoareH87}).
This is used to derive a collection of RHL rules including diagonal and one-side rules
as well as relational \rn{Disj}/\rn{Conj}, for imperative commands including the heap.

Beckert and Ulbrich~\cite{BeckertU18short} survey some of the main ideas in relational verification and describe a
range of applications and works on verification.  
Maillard et al~\cite{MaillardHRM20} introduce a general framework for relational logics,
applicable to a range of computational effects such as exceptions.
Aguirre et al~\cite{AguirreBGGS19} develop a logic based on relational refinement types, for 
terminating higher order functional programs, 
and provide an extensive discussion of work on relational logics.
Recent proceedings of CAV include quite a few papers on relational verification,
and further perspectives can be found in the report from a recent Dagstuhl Seminar on program equivalence~\cite{LahiriMSU18}.

Numerous works develop variations and extensions of the ideas 
in this paper.
Terauchi and Aiken~\cite{TerauchiAiken} observe that sequenced product necessitates use of strong intermediate assertions, and use a dependency type system to guide the construction of more effective products. They also coin the term 2-safety.
Several works focus on modular reasoning and product constructions that enable use of relational specs for procedures~\cite{GodlinS08,ZaksP08,LahiriHKR12,hawblitzelklr13,LahiriMSH13,BanerjeeNN16short,WoodDLE17,EilersMH20}.
Sousa and Dillig~\cite{SousaD16} formulate a logic for $k$-safety, with notation that stands for ``any product'' and may be understood as providing for lazy product construction.
Eilers et al~\cite{EilersMH20} give a $k$-product encoding that lessens code duplication.
Whereas many works handle only lockstep alignment of loops,
some cover the 3-condition automata~\cite{Beringer11,BartheGHS17};  
Shemer et al~\cite{ShemerGSV19} provide for more general alignment 
and infer state-dependent alignment conditions. 
Other works on inferring or expressing effective alignments include~\cite{GirkaEtalPPDP17,PickFG18,ChurchillP0A19}.
Product constructions for $\forall\exists$ properties appear in~\cite{BartheCK13,ClochardMP20}.

Richer formalisms like Dynamic Logic~\cite{DHS05,KeYbook} and embedding in higher order 
logic~\cite{GrimmMFHMPRRSB18,AguirreBGGS19,MaillardHRM20} have their advantages
and can address reasoning like rule \rn{CmdFun} and the linking
of procedures to their implementations which is often left semi-formal.
But such embeddings, in particular, are far from providing the level of automation (and teachability!) 
that more direct implementations of HL/RHL can provide.
Completeness results show how HL/RHL suffice for proving correctness judgments.

\section{Conclusion}\label{sec:concl}

I spelled out a number of patterns of reasoning for program relations and relational properties of programs, in terms of product automata that model pairs of executions, and also as rules of relational program logic.
Almost all the rules can be found in at least one prior publication but some ``obvious'' and useful rules are missing in several papers.
Spelling out the inductive assertion method for relational properties, as Francez~\cite{Francez83} does, 
makes explicit the alignment principles that should be embodied in deductive rules,
guiding the design of such rules.  
On this basis I introduced the notion of alignment completeness,
leaving its formalization to future work; it should be done for a more general form of product 
than the one I chose for expository purposes.

To streamline notation I focused on 2-run properties but there is strong motivation for some 3-run
(e.g., transitivity).
I am not aware of fundamentally different techniques or principles for $k$-run that are not at hand for 2-run.\myendnote{%
   Another topic omitted for lack of space is the soundness of the relational rules under
   $\forall\exists$ interpretations. 
   As in the case of partial versus total correctness interpretation of unary specs,
   most of the rules are sound but loop rules must be changed.
   Different notions of adequacy are needed for product automata.
}

Although several papers have described the need for $k$-safety properties in order 
to reason about unary correctness, to my knowledge this pattern of reasoning has not been 
provided by relational logics (aside from those embedded in expressive higher order logics).
I present a new rule for this (\rn{CmdFun}) that stays within the limited resources of RHL, i.e., assertions, unary correctness, and relational correctness judgments.



A couple of years ago I moved to a smaller office.
While winnowing paper files I came across the paper by Francez, 
which I had acquired but not fully appreciated when working full time at IBM as a programmer in the '80s.  
The dearth of citations shows I am not alone in not finding it when I searched online for relevant work.
My copy is a publisher's reprint, affixed with stickers that indicate IBM paid 
a fee. Such stickers became obsolete but the flow of scientific knowledge is still too tangled with commerce.

\paragraph{Acknowledgments.}

The paper was improved thanks to comments from 
Krzysztof Apt, Anindya Banerjee, Gilles Barthe, Ramana Nagasamudram, and anonymous reviewers.
The research was partially supported by NSF CNS 1718713 and ONR N00014-17-1-2787. 


\iflongver
\begingroup
\parindent 0pt\parskip 2ex
\def\enotesize{\normalsize}\theendnotes
\endgroup
\fi

\bibliographystyle{splncs04}
\bibliography{../../texinputs/new.bib,../../texinputs/naumann.bib}

\end{document}